# Direct volumetric reconstruction for highly compressive x-ray fluorescence ghost tomography

A. Ben-Yehuda[1], A. Rack[2], S. Shwartz[1,*], and N. Viganò[3,*]

[1]*Physics Department and Institute of Nanotechnology and Advanced Materials, Bar Ilan University, Ramat Gan 52900, Israel*

[2]*ESRF — The European Synchrotron, Grenoble 38043, France*

[3]*IRIG-MEM, CEA, Université Grenoble Alpes, Grenoble 38000, France*

*Corresponding author: sharon.shwartz@biu.ac.il

*Corresponding author: nicola.vigano@cea.fr

## Abstract

X-ray fluorescence (XRF) enables element-specific, nondestructive imaging, but conventional raster scanning scales poorly with sample size, particularly for tomography, because measurements must be repeated at every projection angle and spatial position. We demonstrate direct volumetric XRF ghost tomography, which replaces point-by-point acquisition with compressive structured illumination and multiplexed fluorescence detection. Rather than reconstructing projections at each angle and then applying standard tomographic reconstruction, we recover the three-dimensional elemental distribution by solving a single inverse problem that jointly incorporates measurements from all angles. For a volume of 2.8 million voxels, we reconstruct the elemental distribution from only 400 measurements per angle, achieving a 43× reduction relative to raster scanning while maintaining spatial resolution and contrast. By exploiting sparsity directly in the volumetric domain, this approach enables scalable, multi-element XRF tomography of large and heterogeneous samples under stringent acquisition time constraints.

# 1. Introduction

X-ray fluorescence (XRF) microscopy is a cornerstone of quantitative chemical mapping, providing element-specific contrast in two and three dimensions across a broad range of length scales, from tens of nanometers to centimeters [1–9]. It plays a central role in sample characterization in fields ranging from materials science [10] and chemistry [11] to cultural heritage studies [12]. The elemental specificity of XRF stems from core-level electronic transitions: following inner-shell excitation, atoms relax by emitting fluorescent photons at characteristic energies determined by the atomic species. Energy-dispersive detectors resolve these fluorescence lines in the recorded XRF spectra, enabling unambiguous elemental identification. In practice, local XRF spectra, typically acquired using a narrow (pencil-beam) raster scan, are analyzed to reconstruct quantitative spatial maps of chemical composition [13]. In such schemes, each beam position corresponds to a pixel in the reconstructed hyperspectral image.

While conceptually straightforward, this sequential acquisition scales poorly with field-of-view (FoV) and resolution. The bottleneck is amplified in tomography, where a complete raster scan of the entire FoV must be repeated at each projection angle. For complex samples with a large number of voxels, this leads to prohibitively long acquisition times, often incompatible with synchrotron beamtime constraints. Moreover, tight focusing produces a high local dose rate, which can introduce radiation damage, heating, sample deformation, and drift that degrades image quality [14].

Previous work has demonstrated that XRF can be combined with computational ghost imaging (CGI) [15] to substantially reduce the number of measurements required for elemental imaging compared with raster scanning [9,16–20]. In this approach, the sample is illuminated by a sequence of spatially structured intensity patterns that span the FoV [21–25], while a single-pixel, energy-resolving detector records the fluorescence spectrum for each illumination pattern. Elemental maps are then reconstructed computationally from the correlations between the known illumination patterns and the measured signals. For objects exhibiting spatial correlations, CGI enables reconstruction from fewer measurements than pixels by leveraging compressive sensing (CS), which exploits sparsity in a suitable representation [26].

A direct (one-step) 3D reconstruction, introduced in the context of ghost tomography by Kingston and colleagues [27], enforces sparsity in the volumetric domain and jointly incorporates measurements from all projection angles. In this formulation, the object is recovered by solving a single inverse problem, rather than through angle-by-angle projection reconstruction followed by tomographic inversion. Because typical representations are more sparse in the volume than in projection space, this approach enables stronger compression.

Importantly, while direct volumetric reconstruction has been proposed in other ghost tomography contexts [27], its application to x-ray fluorescence is particularly powerful due to the nature of the underlying measurements. In raster-scanning XRF, each measurement probes a localized spatial position at a given projection angle, yielding a well-determined system with little redundancy across measurements. In contrast, structured illumination in CGI encodes global spatial information into each measurement, introducing the degeneracy required for compressive reconstruction. This multiplexing is essential for direct volumetric inversion, as it enables sparsity to be enforced across the full dataset rather than on independently reconstructed projections.

"Here, we implement this framework for x-ray fluorescence ghost tomography (XRF-GT) and demonstrate its experimental realization within a tailored compressive-sensing scheme We recover 3D elemental maps of 164 × 164 × 105 = 2,824,080 voxels from only 400 structured-illumination realizations per angle. With 276 angles, this corresponds to 110,400 total measurements, compared with 4,752,720 required for full angular coverage by pencil-beam scans (43× compression). This approach enables practical XRF tomography of large, heterogeneous samples, for which raster-scan and two-step ghost tomography pipelines become prohibitive even at state-of-the-art facilities.

## 2. Experimental setup

The experimental configuration is shown in Fig. 1. A structured mask mounted on a motorized translation stage is positioned upstream of the sample. For each GI realization, the incident beam illuminates a different region of the mask, generating a distinct structured illumination pattern at the sample plane. The sample is mounted on a rotating stage for tomographic acquisition. Downstream of the sample, a 2D imaging detector records the transmitted intensity, while a silicon drift detector (SDD), placed adjacent to the sample and approximately orthogonal to the incident beam direction, records the XRF signal.

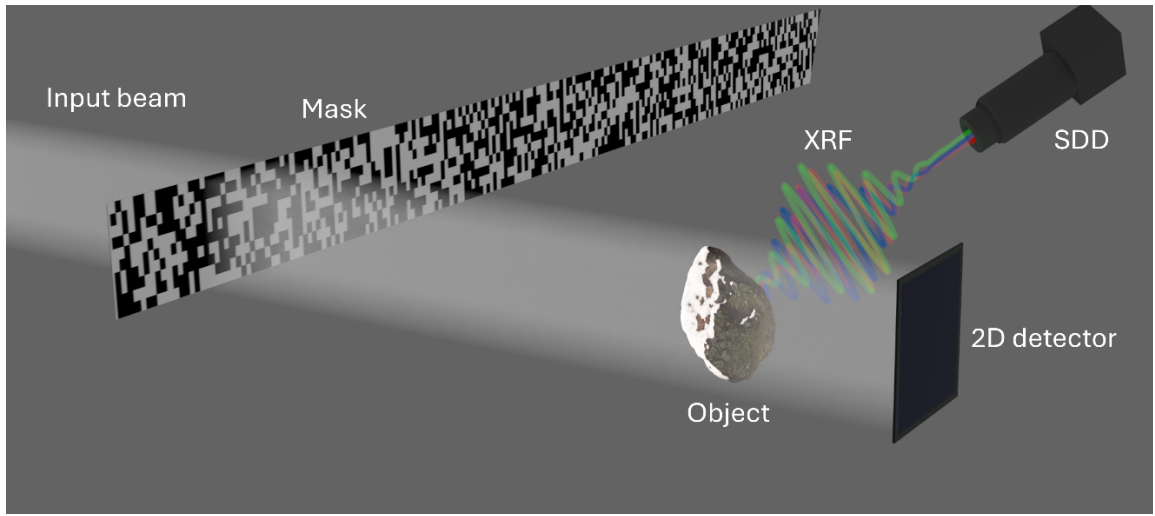

**Fig. 1 Schematic of the experimental setup:** A tungsten mask modulates the incident beam, generating spatially random illumination patterns that irradiate the object. The transmitted intensity is recorded by a 2D detector, while the emitted fluorescence is collected by a silicon drift detector. Both signals are acquired simultaneously as the sample is rotated. Each fluorescence measurement integrates contributions from the entire illuminated region, thereby encoding global spatial information. Element-specific images are reconstructed computationally from correlations between the known illumination patterns and the measured signals, in contrast to raster-scanning XRF, where spatial information is acquired point by point.

The experiment was performed at beamline ID19 of the European Synchrotron Radiation Facility (ESRF) using a broadband incident beam centered at 35 keV with dimensions of 2 mm X 1.2 mm. The sample consisted of Cu wires (40 µm diameter), Zr sheets (20 µm thickness), and spherical Ag particles (maximum particle size 45 µm), enclosed in epoxy resin containing light trace elements.

To generate structured illumination, the mask was translated in 0.06 mm steps, producing 900 distinct realizations. For each mask position, a full tomographic scan was acquired by rotating the sample over 360° in 0.65° increments, yielding 552 projection angles per realization. This angular sampling exceeded that required for the 164 × 164 reconstruction grid. Therefore, the data were binned by a factor of two along the angular dimension, resulting in 276 projection angles per mask realization. To minimize mismatches caused by beam drift and flux fluctuations, the XRF and imaging detectors were operated simultaneously at every angle. The XRF spectra were integrated for 0.1 s per angle [17].

## 3. Data processing and reconstruction

The reconstruction workflows of the direct and conventional two-step GT approaches are illustrated in Fig. 2. In the two-step XRF-GT scheme (Fig. 2(a)), element-specific 2D ghost projections are first reconstructed independently for each rotation angle, followed by a standard tomographic inversion to obtain the 3D elemental distribution. In contrast, the direct XRF-GT (Fig. 2b) recovers the 3D elemental volume directly from the raw measurements by solving a single three-dimensional inverse problem.

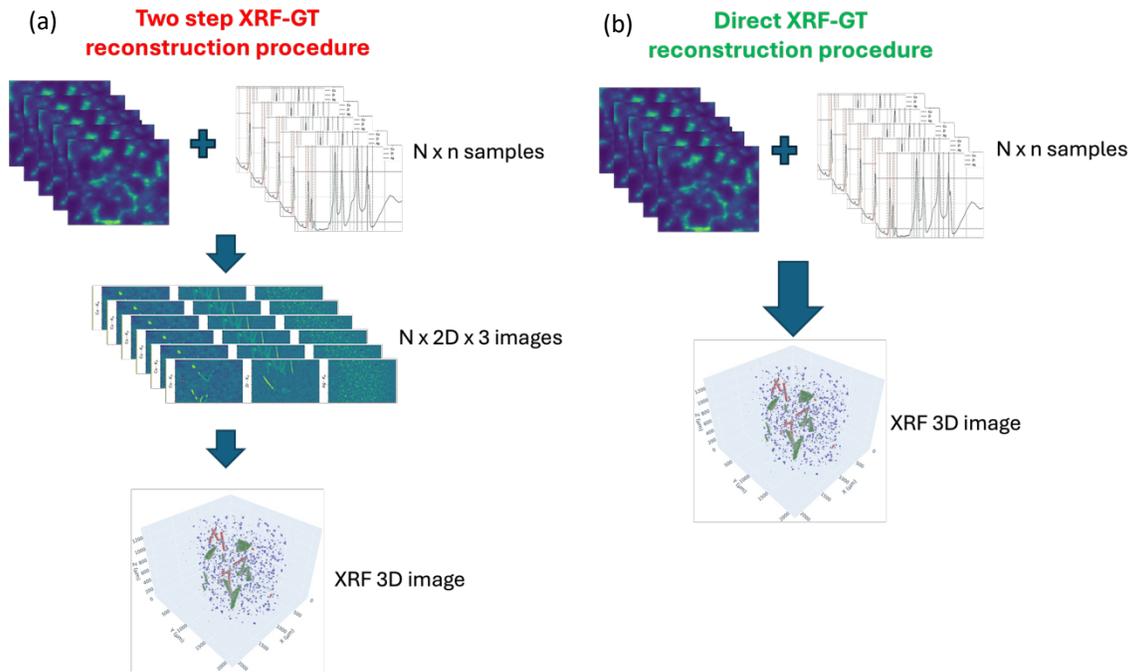

**Fig. 2. Comparison between conventional two-step GT and direct volumetric reconstruction in XRF-GT:** In the standard approach, angle-resolved ghost projections are first reconstructed independently and subsequently combined via tomographic inversion. In contrast, the direct approach recovers the three-dimensional elemental distribution by solving a single inverse problem that jointly incorporates measurements from all projection angles. This formulation leverages the multiplexed nature of structured illumination in CGI, where each measurement encodes global spatial information. This inherent redundancy enables compressive reconstruction and allows sparsity to be enforced directly in the volumetric domain, in contrast to raster-scanning XRF, where measurements are localized and do not provide such degeneracy.

The two approaches differ in how sparsity is enforced: the two-step pipeline reconstructs projections independently at each angle, whereas the direct formulation imposes sparsity directly in the volumetric domain and jointly incorporates all measurements across masks and projection angles. The direct reconstruction is formulated as the minimization problem:

$$\hat{x} = \operatorname*{argmin}_{x} \left\{ \frac{1}{2} \|MRx - b\|_2^2 + \lambda \|Hx\|_1 \right\} \qquad (Eq.\,1)$$

where $\hat{x}$ is the reconstructed elemental distribution, $R$ is the Radon transform preceding the CGI projection operator $M$, $b$ is the vector of measured fluorescence signals, and $\lambda$ controls the influence of the sparsity prior. The operator $H$ corresponds to the point-wise modulus of the gradient,

$$H(\cdot) := |\nabla(\cdot)| : \mathbb{R}^3 \to \mathbb{R}, \qquad (Eq.\,2)$$

which implements total-variation (TV) regularization. A detailed derivation of Eq. (1) is presented in Sec. S6, and an intuitive example of sparsity behavior is given in Sec. S7 of the supplementary material. The regularization weight $\lambda$ is selected via cross-validation to balance sparsity enforcement and preservation of the fine structural details.

## 4. Results

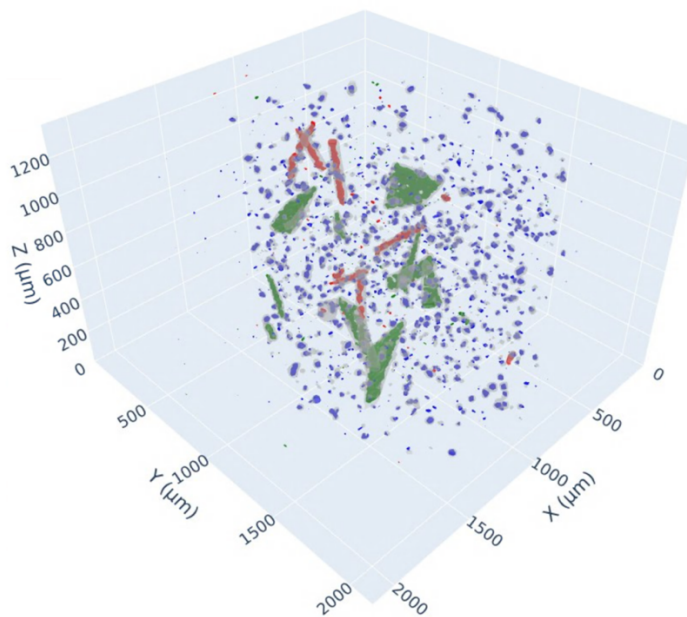

**Fig. 3 Isosurface:** XRF Isosurface of the Cu (red), Zr (green) and Ag (blue) reconstructed by direct XRF-GT from 400 structured-illumination measurements per angle over the absorption isosurface (gray).

We demonstrate that direct XRF-GT reconstructs a 3D elemental map of the sample using only 400 structured-illumination measurements per angle. By comparison, an equivalent pencil-beam XRF scan at the same voxel grid would require 17,220 (105x164) beam positions per angle, corresponding to a 43x reduction in measurements. The isosurface rendering in Fig. 3 reveals Cu wires, Zr foils, and Ag particles at their expected locations, with no visible streaking or structured reconstruction artifacts.

To validate our approach, we compare direct reconstructions with results from the conventional two-step XRF-GT pipeline, and with an independent transmission tomogram (Fig. 4). Fig. 4(a)-(c) shows representative tomographic slices. With 400 realizations per angle, only the direct reconstruction reveals well-defined features that can be readily compared with the transmission slice. In contrast, the two-step reconstruction shows pronounced background fluctuations and circular artifacts that are absent in the direct reconstruction. The distinct morphologies of the Cu wires, Zr foils, and Ag particle enable their identification in the transmission tomogram, and the spatial correspondence with the direct XRF reconstruction confirms the fidelity of the recovered elemental distributions.

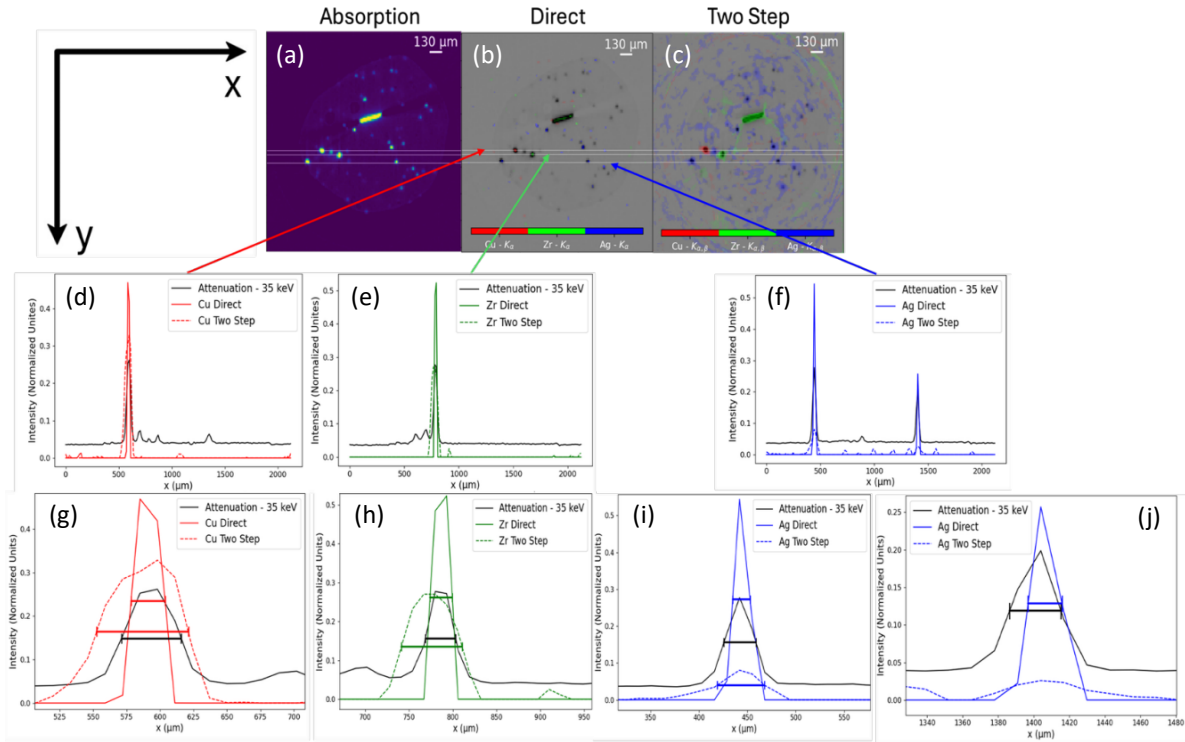

**Fig. 4. Tomograms and cross sections:** (a) Transmission tomogram, XRF tomograms reconstructed from 400 measurements per angle (b) direct and (c) two step. (d-f) Corresponding cross sections at y=1131 µm, 1170 µm and 1248 µm for Cu, Zr and Ag, respectively at z=975 µm . Zoomed-in views of the regions indicated by arrows in (a–f): Cu and Zr in (g,h) and Ag in (i,j), where Ag exhibits two peaks.

We further quantify this agreement using line profiles extracted from identical cross-sections as shown in Fig. 4(d)-(f) . For Cu, Zr and Ag, the direct reconstruction produces profiles with a full width at half maximum (FWHM) of approximately 30 µm, 30 µm, and 20 µm. The corresponding values obtained with the two-step reconstruction are about 70 µm, 70 µm, and 50 µm, respectively.

The transmission tomograms yield 44 μm, 35 μm, and 33 μm. The cross-sections also reveal substantially lower background levels for the direct approach than the two-step pipeline. These differences are most pronounced in low signal regions, where the two-step approach exhibits elevated background and reduced contrast. The broader FWHM obtained with the two-step reconstruction reflects its lower image quality and reduced ability to recover fine structural features compared with the direct approach. The remaining differences relative to the transmission data arise primarily from the higher spatial resolution of the transmission tomogram, which is limited by the detector pixel size, whereas the CGI resolution is additionally constrained by the mask feature size. Small discrepancies between the fluorescence and transmission profiles may also arise from physical effects such as self-absorption in the fluorescence signal.

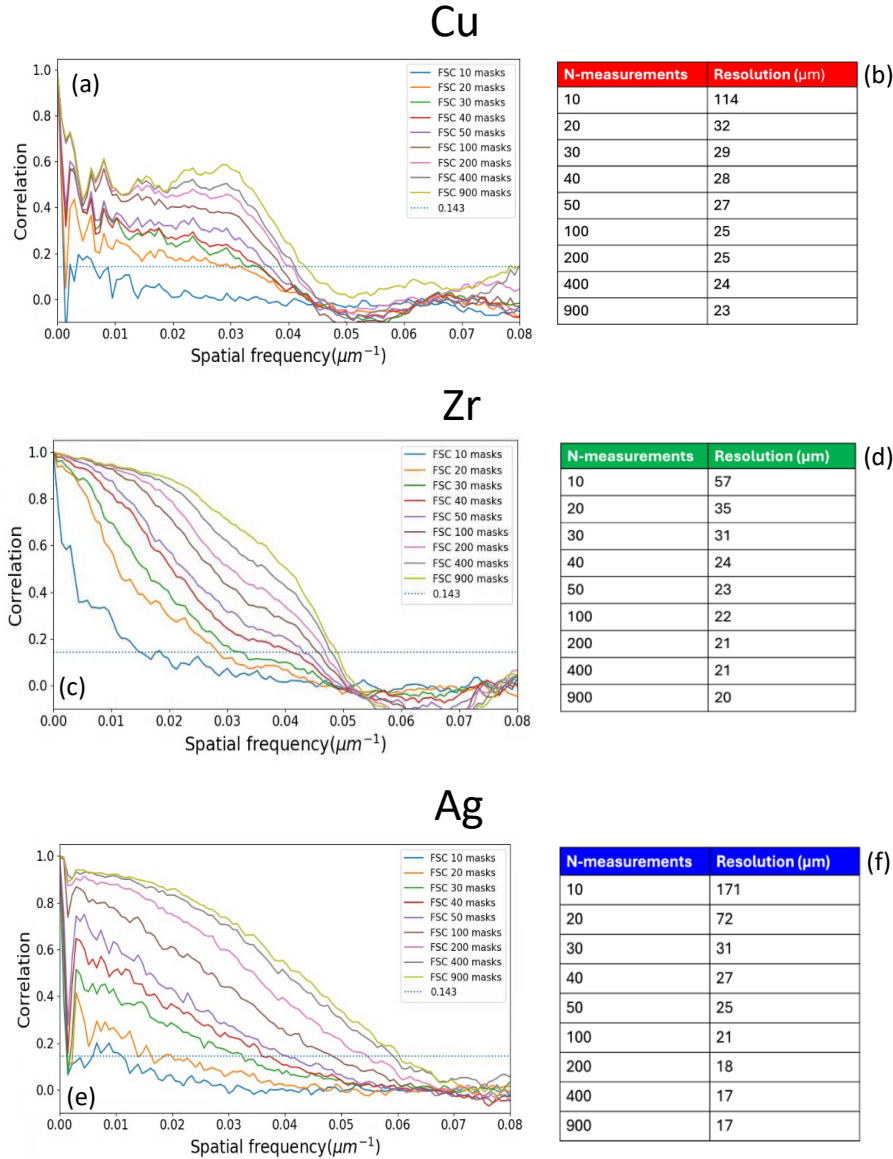

**Fig. 5. Spatial resolution quantification using Fourier shell correlation (FSC):** FSC curves and the resolution of Cu (a), (b), Zr (c), (d) and Ag (e), (f) in accordance with the number of measurements.

To characterize the performance of the direct approach, we estimate the spatial resolution using Fourier shell correlation (FSC) [28]. Fig. 5 shows FSC curves for the Cu, Zr and Ag reconstructions for increasing numbers of masks. For all elements, the curves extend to higher spatial frequencies as the number of masks increases. Using the 0.143 correlation criterion [29], the corresponding resolution estimates are summarized as a function of measurement number in Fig. 5. With 400 masks, the achieved resolutions for Cu, Zr and Ag are 24 μm, 21 μm, and 17 μm, respectively. These values are comparable to the results obtained with 900 masks, differing only slightly in the cutoff spatial frequency. For fewer than 200 masks, the FSC curves show a pronounced loss of high-frequency information.

## 5. Discussion and conclusions

We have demonstrated XRF-GT using a direct reconstruction framework that jointly inverts measurements across projection angles to recover 3D multi-element volumes under strongly undersampled conditions. Using only 400 structured-illumination measurements per angle, we reconstructed elemental distributions comprising 2,824,080 voxels and achieved spatial resolutions of 24 μm (Cu), 21 μm (Zr), and 17 μm (Ag). Compared with pencil-beam scanning-based XRF tomography, this corresponds to an effective 43× per-angle compression. More generally, this work illustrates how multiplexed illumination combined with direct volumetric inversion can substantially reduce the measurement burden of XRF tomography.

More broadly, the proposed approach changes how acquisition scales with measurement number. In conventional raster-scan XRF tomography, the acquisition time is determined by the dwell time required to accumulate sufficient fluorescence counts at each spatial position, combined with the need to repeat this sampling at every projection angle. Direct XRF-GT replaces sequential point sampling with illumination that encodes global information across the FoV. By exploiting sparsity directly in the 3D elemental distribution rather than in intermediate projections, our method substantially reduces the number of measurements required per angle. This reduction enables multi-element 3D XRF tomography within practical synchrotron beamtime and dose constraints, and facilitates studies of large heterogeneous specimens, dose-sensitive materials, and experiments requiring repeated measurements.

While direct XRF-GT offers substantial gains in measurement efficiency, several practical considerations remain. Reconstruction quality benefits from sparsity or compressibility of the 3D elemental distribution, and samples with more complex or nonuniform elemental distributions may require increased sampling. In addition, the joint 3D inversion increases computational and memory requirements, and the resulting data volumes can make straightforward application of machine-learning approaches [30] challenging at present. Looking forward, improved physics-based modeling, faster solvers, and adaptive illumination strategies may further extend XRF-GT toward higher resolution, larger volumes, and lower-dose operation, enabling studies of more complex samples within shorter acquisition times and lower dose budgets.

**Data Availability.** The data that support the findings of this study are available from the corresponding author upon reasonable request.

**Acknowledgments.** This work was supported by the Israel Science Foundation (ISF), Grant No. 2208/24. We acknowledge the European Synchrotron Radiation Facility (ESRF) for provision of synchrotron radiation facilities under proposal ID MI-1505.

# Supplementary Materials

This supplementary material provides additional details on the reconstruction framework, sparsity analysis, and experimental procedures used in the direct XRF GT study presented in the main text. It includes descriptions of the forward models for ghost tomography, comparisons between two-step and direct reconstruction strategies, analyses of sparsity in different signal representations, and supporting validation of reconstruction performance.

**1. Convergence Factor Calculation**

The Convergence Factor (CF) is used to evaluate the agreement between reconstructed elemental maps and reference images. It is calculated as the Mean Squared Error (MSE) between the manual classification of the absorption image and the XRF-GT reconstruction, normalized by the mean pixel intensity of the reference image to avoid scaling effects.

$$CF = \frac{MSE(Manual, Reconstructed)}{Mean(pixels\ in\ reference\ image)}, \qquad (Eq.\ S1)$$

where:

- *MSE* - mean Squared Error.
- *Manual* - manually classified elemental regions derived from the absorption image.
- *Reconstructed* - image obtained from the XRF-GT reconstruction.

Fig. S1 shows the convergence factor dependence on the number of masks (measurements) used for each element (Cu, Zr, Ag).

## 2. Reconstruction of Absorption and XRF Tomograms

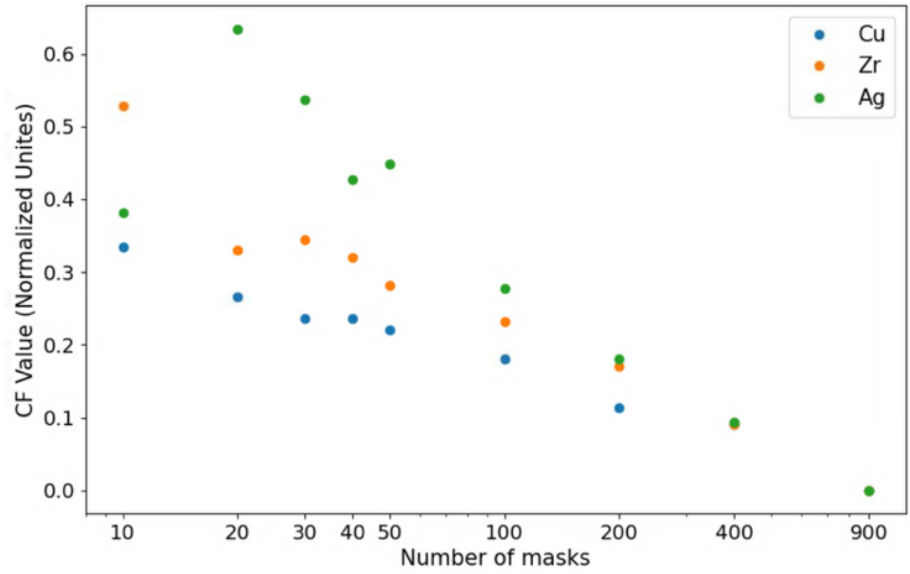

**Fig. S1 Convergence Factor as a function of the number of masks (measurements) for each Cu, Zr and Ag.**

Fig. S2 illustrates the improved reconstruction quality achieved using the direct XRF-GT approach for low measurement counts (50 measurements per angle). It compares the direct (Fig. S2(a)) and the two-step (Fig. S2(b)) methods. The corresponding cross-sectional maps for Cu, Zr, and Ag, shown in Fig. S2(c–e), further emphasize these improvements, particularly through the substantial reduction of background noise and the clearer delineation of elemental structures.

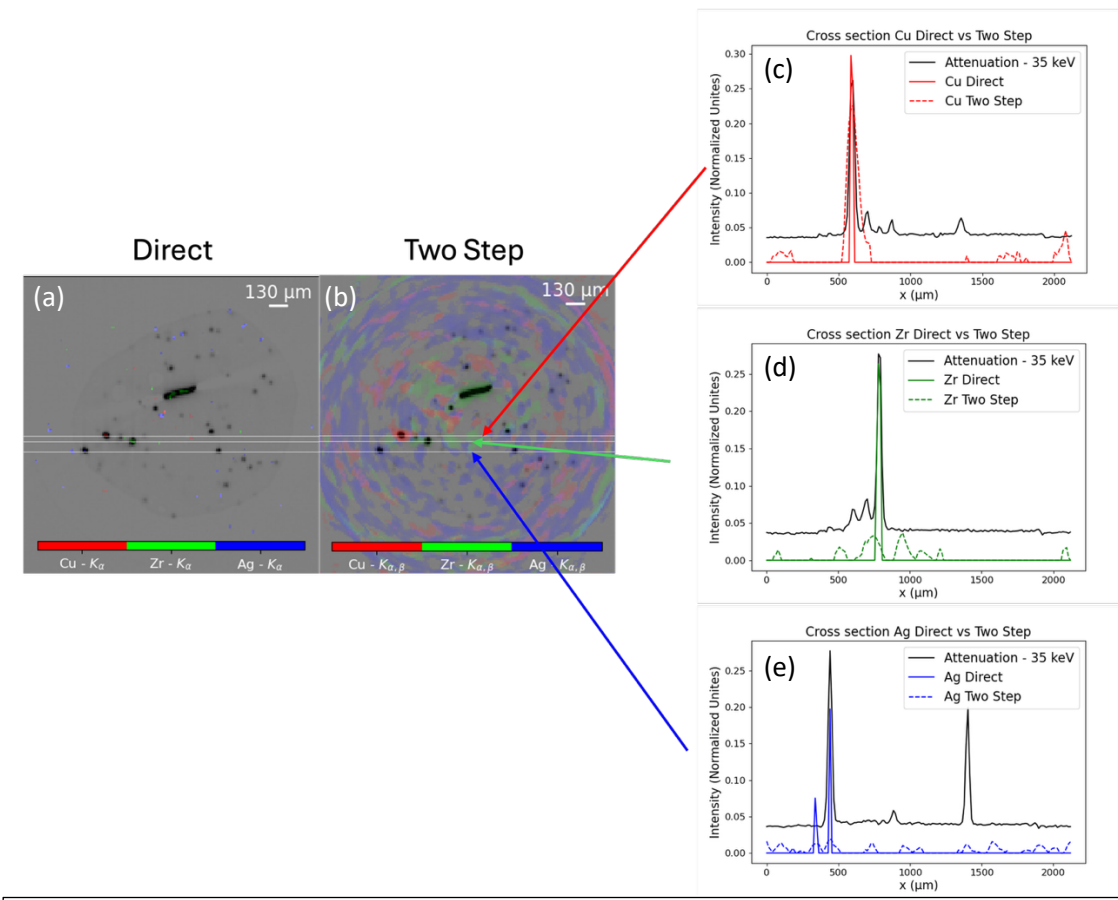

**Fig. S2. Absorption and XRF Tomograms and their cross sections for low measurement count (50 measurements):** (a) Direct XRF-GT tomogram, (b) Two step XRF-GT tomogram, (c-e) cross sections of Cu, Zr and Ag for Absorption, direct XRF-GT and two step XRF-GT.

## 3. Spectral Measurement and Element Identification

The measured fluorescence spectrum of the sample is shown in Fig. S3. Each spectrum contains characteristic energy peaks corresponding to specific elements present in the sample. In this study, Cu, Zr, and Ag were identified. These spectra were analyzed to generate calibrated elemental maps that serve as input for the 3D tomographic reconstruction.

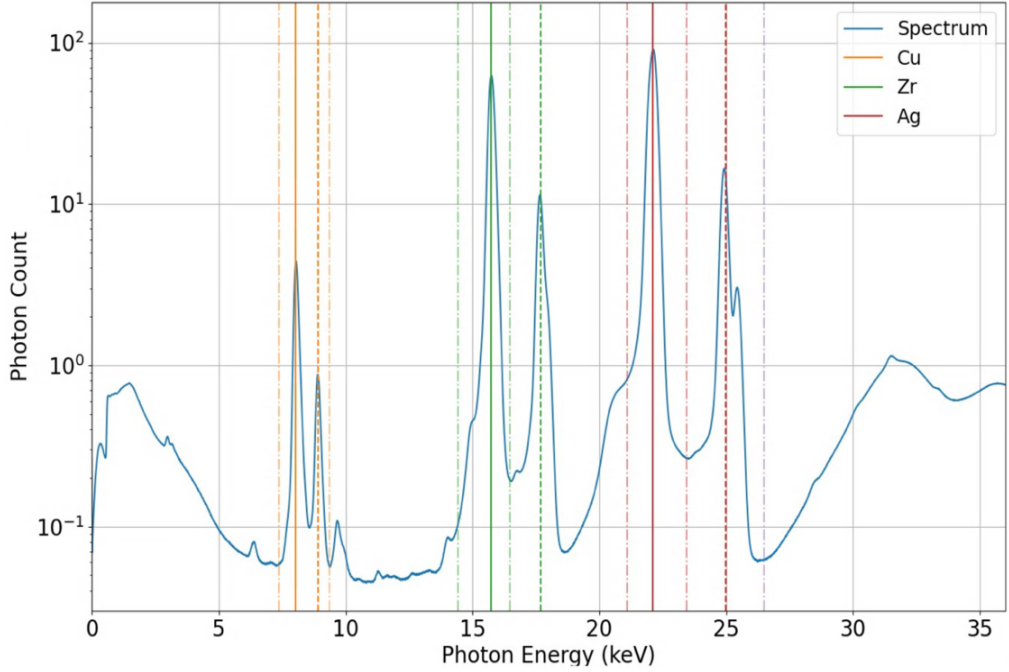

**Fig. S3. Measured spectrum of the sample.**

**4. Experiment details**

Fluorescence spectra were collected using a Hitachi Vortex 90EX detector controlled by a XIA FalconX module. The imaging system consisted of a so-called Hasselblad system, with two identical lenses (100 mm focal length) in tandem configuration (giving ~ x1 magnification), with a 500 µm LuAG:Ce scintillator and a "pco.edge 5.5" camera. The detector assembly was positioned approximately 5 cm downstream of the sample, with an effective pixel size of 6.5 µm.

**5. Illumination patterns decoupling**

To isolate the illumination patterns, we extract the masks from the transmission data following the approach of Manni et al. [1]. We form an angle-summed projection by integrating the transmission images over all rotation angles. This procedure suppresses high-frequency features from the rotating sample while preserving the static high-frequency structure of the mask. The resulting contrast separation enables an improved estimate of the mask patterns used in the CGI reconstruction.

**6. Reconstruction of Sparse Signals**

We denote $\hat{x}$ as the reconstructed signal. The reconstruction problem can be formulated as:

$$\hat{x} = \underset{x}{\mathrm{argmin}} \left\{ \frac{1}{2} \|Ax - b\|_2^2 + \lambda \|Hx\|_1 \right\} \qquad (Eq. S3)$$

where A is the measurement operator, b is the vector of measurements, λ controls the strength of the sparsity prior and H [2] is the decomposition basis while $l_1$-norm is non-smooth, but convex. Eq. S3 can be minimized by algorithms that accept the use of convex non-smooth terms in the objective function [3,4].

## 7. Computational Ghost Imaging and Tomography

For CGI, the measurement operator A corresponds to the projection matrix M, whose rows contain the flattened illumination patterns measured at the sample plane. Under common structured illumination schemes (e.g., random Bernoulli or deterministic Hadamard patterns), the conditions allowing the substitution of the $l_0$-norm with the convex $l_1$-norm are typically satisfied.

For GT, two reconstruction strategies can be considered:

**2-step reconstruction.** In the conventional approach, ghost projections are reconstructed independently at each projection angle:

$$\hat{p} = \underset{p}{\mathrm{argmin}} \left\{ \frac{1}{2} \|Mp - b\|_2^2 + \lambda \|Hp\|_1 \right\}, \qquad (Eq. S4a)$$

where p is the stack of reconstructed projections.
The volumetric distribution is then obtained through a tomographic inversion,

$$\hat{x} = \underset{x}{\mathrm{argmin}} \left\{ \frac{1}{2} \|Rx - \hat{p}\|_2^2 \right\}, \qquad (Eq. S4b)$$

with R representing the Radon transform.

**Direct volumetric reconstruction.** Alternatively, the tomographic reconstruction can be formulated as a single inverse problem by incorporating the Radon transform directly into the forward model:

$$\hat{x} = \underset{x}{\mathrm{argmin}} \left\{ \frac{1}{2} \|MRx - b\|_2^2 + \lambda \|Hx\|_1 \right\}, \qquad (Eq. S5)$$

In this formulation, the 3D elemental distribution x is recovered directly from the GT measurements without intermediate projection reconstruction. In our implementation, the operator H corresponds to the magnitude of the spatial gradient $H(\cdot) := |\nabla(\cdot)| : \mathbb{R}^3 \to \mathbb{R}$, which implements total variation (TV) regularization.

## 8. Sparse Representations in Tomography

To illustrate the sparsity differences between reconstruction and projection domains, we consider a simple tomographic example based on the Shepp–Logan phantom. By comparing different representations of the phantom and its corresponding sinogram, we evaluate how sparsity varies between the reconstruction space and the acquisition space. This comparison provides insight into why enforcing sparsity directly in the reconstruction domain can be advantageous for tomographic reconstruction.

Fig. S4 shows different representations of the phantom: the image itself and its spatial gradients. In the gradient representation, homogeneous regions are mapped to zero, leaving only the object boundaries. As a result, the gradient representation is significantly sparser than the image representation.

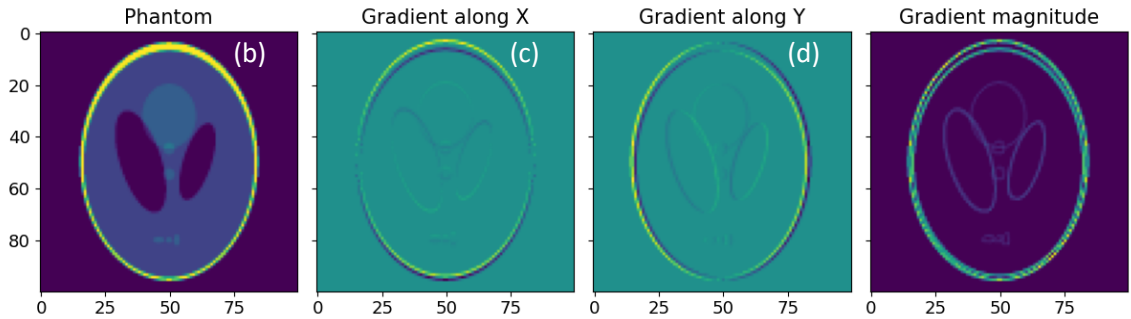

**Fig. S4. Phantom representations:** (a) reconstruction domain, (b) gradient along X, (c) gradient along Y, and (d) magnitude of the gradient vector.

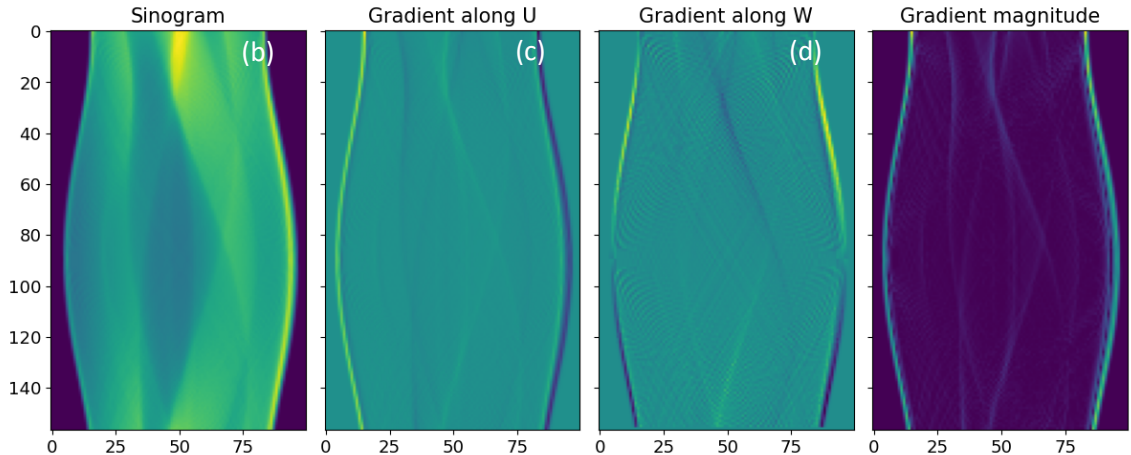

**Fig. S5. Sinogram representations:** (a) acquisition domain, (b) gradient along U, (c) gradient along W, and (d) magnitude of the gradient vector.

Fig. S5 presents the corresponding sinogram representations. In contrast to the reconstruction domain, the sinogram remains relatively dense even in its gradient representation. This difference is quantified using the $l_0$-norm and $l_1$-norm of the corresponding representations, which are presented in Tab. S1:

|  | Phantom | | | Sinogram | | |
| --- | --- | --- | --- | --- | --- | --- |
|  | Image | Gradient (XY) | Gradient (Magnitude) | Image | Gradient (U) | Gradient (Magnitude) |
| $l_1$-norm | 1231.589 | 597.3985 | 20.31548 | 193365.7 | 11050.65 | 427.1116 |
| $l_0$-norm | 4412 | 2310 | 1457 | 13393 | 13549 | 13564 |

**Tab. S1. Norms of different representations in the reconstruction and acquisition domains**

Both norms are significantly smaller in the reconstruction domain than the in the sinogram domain. Moreover, the gradient magnitude representation of the phantom is sparser than the original image representation.

These observations indicate that sparsity is more effectively enforced in the gradient of the reconstruction domain than in projection space. Consequently, reconstruction strategies that impose sparsity directly on the volumetric distribution, such as the direct reconstruction approach used in this work, can achieve higher compression than methods based on intermediate projection reconstruction.